\documentclass{emulateapj}

\shorttitle{Spectroscopy of HR 8799 c}
\shortauthors{Janson et al.}

\begin{document}


\title{Spatially resolved spectroscopy of the exoplanet HR 8799 c\altaffilmark{*}}


\author{M. Janson\altaffilmark{1,4}, C. Bergfors\altaffilmark{2}, M. Goto\altaffilmark{2}, W. Brandner\altaffilmark{2}, D. Lafreni\`ere\altaffilmark{3}}
\email{janson@astro.utoronto.ca}


\altaffiltext{*}{Based on observations collected at the European Southern Observatory, Chile (ESO No.\ 084.C-0072) and at the Subaru telescope, which is operated by the National Astronomical Observatory of Japan.}
\altaffiltext{1}{University of Toronto, Toronto, Canada}
\altaffiltext{2}{Max-Planck-Institute for Astronomy, Heidelberg, Germany}
\altaffiltext{3}{University of Montreal, Montreal, Canada}
\altaffiltext{4}{Reinhardt fellow}


\begin{abstract}
HR 8799 is a multi-planet system detected in direct imaging, with three companions known so far. Here, we present spatially resolved VLT/NACO 3.88--4.10 $\mu$m spectroscopy of the middle planet, HR 8799 c, which has an estimated mass of $\sim$10 $M_{\rm Jup}$, temperature of $\sim$1100 K and projected separation of 38 AU. The spectrum shows some differences in the continuum from existing theoretical models, particularly longwards of 4$\mu$m, implying that detailed cloud structure or non-equilibrium conditions may play an important role in the physics of young exoplanetary atmospheres. 
\end{abstract}


\keywords{planetary systems --- techniques: spectroscopic}



\section{Introduction}

Spectroscopy of extrasolar planets will be a key tool towards understanding their formation and atmospheres, and could eventually lead to identification of chemical biomarkers (e.g. Kaltenegger et al. 2007), thus addressing the question of life elsewhere in the universe. However, the large brightness contrast and small angular separation between a star and planet makes such spectroscopy technologically challenging. In the special geometric case where the orbital plane of the planet is aligned along our line of sight, a spectroscopic signal can be temporally resolved by comparing the combined star and planet light in and out of transit or secondary eclipse. This has been successfully done with the Spitzer and Hubble space telescopes (e.g. Grillmair et al. 2007; Swain et al. 2009). However, the vast majority of exoplanets do not share this fortunate geometry. For these planets, it is necessary to spatially resolve the planetary spectral signal from the stellar one. 

During the past year, several planet candidates from direct imaging have been reported (e.g. Lafreniere et al. 2008, Kalas et al. 2008, Lagrange et al. 2009). Of particular interest here is the HR 8799 planetary system (Marois et al. 2008), where the three companions have been confirmed through common proper motion, orbital motion, and detection of thermal radiation in several photometric bands. This system is in many ways reminiscent of a scaled-up version of our own Solar system. HR 8799 itself is an A5 star of 1.5 $M_{\rm Sun}$. HR 8799 b, c, and d have estimated masses of 7 $M_{\rm Jup}$, 10 $M_{\rm Jup}$, and 10 $M_{\rm Jup}$ and separations of 68 AU, 38 AU and 24 AU, respectively. The mass estimates are based on an age estimate of 60 Myr for HR 8799. Although much higher ages are difficult to exclude categorically, several age indicators (such as the massive debris disk) and arguments from dynamical stability mutually indicate such a young age (e.g. Reidemeister et al. 2009). Modeling of infrared excess by Spitzer (e.g. Chen et al. 2009) has shown that a Kuiper-analogous debris belt exists outside of HR 8799 b, and a debris belt analogous to the Asteroid belt exists inside of HR 8799 d. The planets appear to be largely co-planar with an orbital plane close to face-on, and have low eccentricities.

\section{Observations and Data Analysis}

Spectroscopic runs were performed with IRCS at Subaru and NACO at the VLT during the second half of 2009 with the purpose of detecting the spectroscopic L$^{\prime}$-band signature of the planets, in particular HR 8799 b and c. The IRCS run was aimed primarily at HR 8799 b, and proved the feasibility of the technique, but the thermal background was much higher than predicted, probably due to dust on the image rotator or the high ambient humidity during the run. Hence, no planet spectroscopy could be extracted from the IRCS data. For the NACO run, one out of four half-nights was clouded out and two had thin cirrus clouds. During one of the nights with cirrus, HR 8799 b was included in the slit. Hence, some NACO data of HR 8799 b exists, but given the ambient conditions we focused mainly on getting a strong detection on HR 8799 c during the rest of the time. Hence, the available data on HR 8799 b is insufficient to make a signficiant spectral detection. The HR 8799 c spectroscopy is the topic of this Letter.

The technique employed was slit spectroscopy in the L$^{\prime}$-band range. This choice is based on the combination of facts that the star-to-planet contrast is more favorable there (9.5 mag for HR 8799 c, as compared to 11.6 in H-band), and that an exceptional adaptive optics performance can be achieved in this range (Strehl ratio upwards of 85 \% at 4 $\mu$m, see Janson et al. 2008). For these reasons, a planet like HR 8799 c dominates the stellar PSF at its angular separation. A disadvantage of L$^{\prime}$-band with respect to shorter wavelengths is the higher thermal background, but previous observations had shown that this was not a major limitation. The widest NACO slit of 172 mas was used, which allows for the companion PSF to be entirely included in the slit, and along with the high PSF stability, allows to ensure that PSF chromaticity will have a negligible effect over the relevant wavelength range. The spectral resolution, limited by the FWHM, was approximately 900. The pixel scale was 27 mas/pixel. Individual integration times were 5 sec, and 20 coadds per frame were taken. After excluding bad frames, 102 frames were used for the target characterization (out of 110 taken, i.e. 7\% excluded) and 94 for the PSF characterization (out of 110 taken, i,e, 15\% excluded), giving total integration times of 10200 sec and 9400 sec, respectively.

\begin{figure}
\plotone{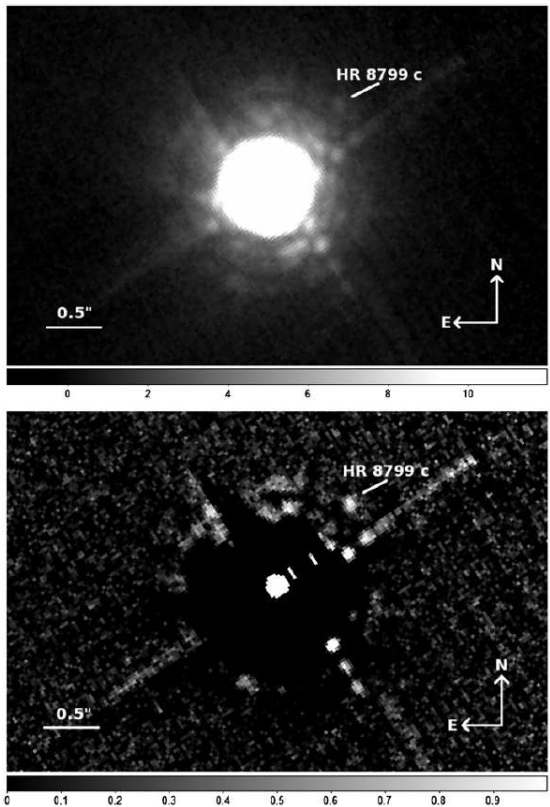}
\caption{Acquisition image of HR 8799. Top: A 150 second L$^{\prime}$-band image of HR 8799 with no high-contrast techniques applied. HR 8799 c can be spotted even in this very short exposure, demonstrating the advantage of this wavelength range for exoplanet spectroscopy. Bottom: The same image with high-pass filtering applied.\label{hrimg}}
\end{figure}

A quick-look L$^{\prime}$-band image of the HR 8799 system was taken in the beginning of the night to optimize the slit alignment. A 150 second sequence without any high-contrast technique applied showed the presence of the planet, confirming that HR 8799 c indeed dominates the stellar PSF at its separation (see Fig. \ref{hrimg}). The slit was aligned along the star-planet axis to ensure an adequate slit positioning and a good PSF characterization of the star, as well as provide an optimal correction of telluric and instrumental transmission effects. The price to pay for this is some ghost images on the detector from internal reflections. Pairwise nodding with 5" nod size was applied for background subtraction. Observations were also taken with a slit rotation of 180$^{\rm o}$ for PSF reference subtraction. Flat field frames were produced with the slit and grism in, pointing at the thermal sky.

\begin{figure}
\plotone{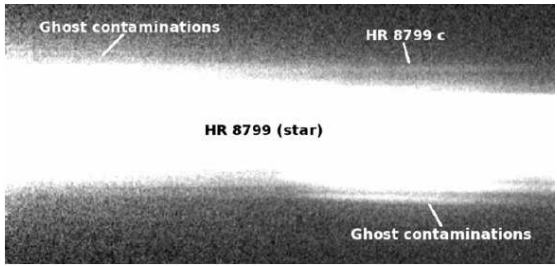}
\caption{Image with the HR 8799 c spectrum before extraction. The spatial direction along the star-planet axis is vertical. The spectral direction is horizontal, with wavelength increasing from left to right.\label{hrorig}}
\end{figure}

The data reduction was performed with dedicated procedures written in IDL. The flat field frame reproduced the expected vertical fringes that occur in 4 $\mu$m spectroscopy. The fringe pattern was constant but shifted laterally with time, so the flat frame had to be digitally shifted to each science frame. Since doing so would eliminate pixel-to-pixel variation correction on the detector, both the image and flat frames first had to be corrected for such variations. This was done with an imaging L$^{\prime}$-band flat which was also acquired during the night. The spectral dependency of the pixel response was negligible, and so this correction could be done satisfactorily. The shift of the spectroscopic flat was determined individually for each science frame by cross-correlation of the fringe pattern with minimization of the residuals. The flat- and dark-corrected frames were pairwise subtracted to remove the thermal background. The spectrum of the star was traced to an absolute center using a Gaussian fit on the PSF core individually for each wavelength, and relative centering was checked using cross-correlation on the core with minimization of the residuals. A trace was then fitted over all wavelengths with 21 pixel median box filtering, and all spectra were shifted to a common center. A collapse of all the frames revealed the spectrum of the planet, at the expected position and with the expected average brightness (see Fig. \ref{hrorig}). The short-wavelength tail of the planetary spectrum interfered with a residual ghost feature, which could not be entirely removed since the PSF reference frames were not taken at identical dithering positions as the science frames. Hence, we count this range as being of insufficient quality and use only the $>3.88$ $\mu$m range where the planetary spectrum is clean.

The star and planet spectra were extracted using 4-pixel apertures (108 mas, close to the FWHM of the PSF). The background for the planet was estimated on the corresponding flux in the PSF reference frame, averaged over 7 neighboring pixels in the wavelength direction so as not to increase the background noise. The background residual spectrum (non-averaged over wavelength) was also used to estimate the error by taking the standard deviation over 21 neighboring pixels in the wavelength direction. Wavelength calibration was achieved using telluric methane lines in the stellar spectrum. Flux calibration was done by first calculating the planet-to-star contrast at each wavelength, which also automatically corrected for telluric and instrumental transmission effects. The absolute scaling was then determined by calculating the blackbody radiation of HR 8799, based on the radius (1.34 $R_{\rm Sun}$) and temperature (7430 K) given in Gray \& Kaye (1999). Although HR 8799 is known to exhibit infrared excess, the corresponding dust has been located at around 8 AU and beyond (Chen et al. 2009), and should be close to face-on. Our photometric aperture of HR 8799 corresponds to a radius of about 0.2 AU, hence the dust should be resolved out and a photospheric blackbody spectrum should therefore be a good representation of what we measure in our data.

\section{Results and Discussion}

\begin{figure}
\plotone{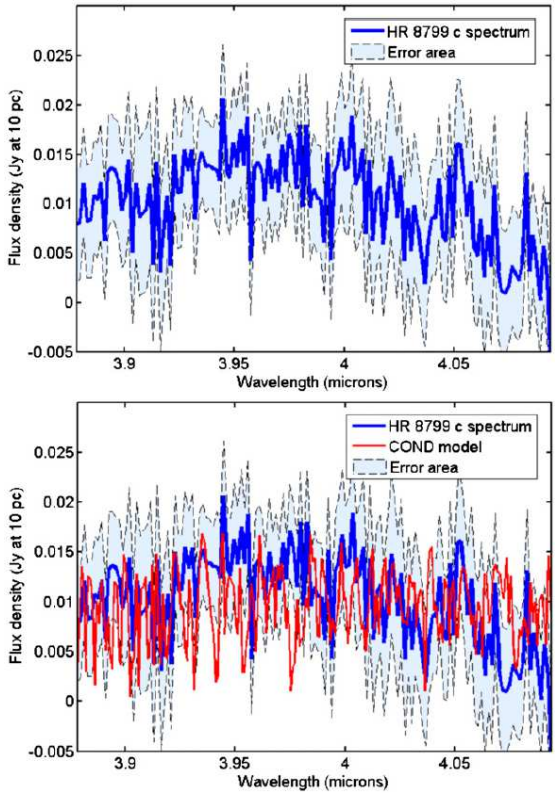}
\caption{Upper: Spectrum of HR 8799 c. The dashed lines and faintly shaded area (light blue in the online version) denote the errors. Lower: Same figure but with a COND model spectrum overplotted as a thinner line (red in the online version).\label{hrspec}}
\end{figure}

\begin{figure}
\plotone{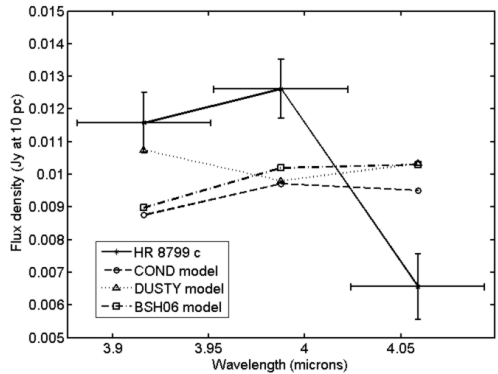}
\caption{The HR 8799 c spectrum (stars with error bars, solid line) undersampled into three broad bins to illustrate the large-scale slope. Also plotted are COND (circles, dashed line),  DUSTY (triangles, dotted line) and BSH06 (squares, dash-dotted line) model spectra at the same sampling.\label{hrbin}}
\end{figure}

The spectrum of HR 8799 c is shown in Fig. \ref{hrspec}. Also plotted in an adjacent panel is a COND model spectrum (Baraffe et al. 2003) with $T_{\rm eff} = 1100$ K, $\log g = 4.0$ and $R_{\rm pl} = 1.3$ $R_{\rm Jup}$. The model can readily reproduce the mean brightness of the spectrum within the planetary parameters set in Marois et al. (2008) based on the bolometric luminosity and cooling models. However, the slope is different with the planetary spectrum showing a peak around 4 $\mu$m and the model spectrum being largely flat. This cannot be reproduced by changing any of the parameters in the COND model. Changing the temperature essentially only changes the mean flux of the spectrum, and the $\log g$ parameter has very little impact at all, aside from affecting the radius for a given mass, which again can only change the mean flux. The difference is more easily seen in Fig. \ref{hrbin}, where the data have been sub-sampled into three broad bins. There is a slope of 5.5$\sigma$ signficiance in the long wavelength end, and likewise a deviation of the same order from the model. 

Hence, there are features in the spectrum that the COND model cannot explain. Part of the problem is probably dust in the atmosphere -- COND assumes dust-free conditions. However, a DUSTY model (Chabrier et al. 2000) with the same range of parameters can also not reproduce the spectrum (see Fig. \ref{hrbin}). Models with intermediate abundances of dust have also been unable to reproduce the full set of observed properties of the HR 8799 planets (Marois et al. 2008). We have also tried the models of Burrows et al. (2006) which we refer to here as BSH06, but the results are very similar to those of COND (also plotted in Fig. \ref{hrbin} with the same parameters, except that $\log g = 4.5$ in this case). The results therefore imply that a more detailed treatment of dust in the models is necessary -- or perhaps, that non-equilibrium chemistry is involved (Fortney et al. 2008). Non-equilibrium models are worth to explore as they predict large differences in the spectrum as function of metallicity. In particular, a broad dip in the spectrum caused by an increased presence of CO leads to depression of the flux longwards of 4 $\mu$m for high-metallicity non-equilibrium atmospheres in the Fortney et al. (2008) models. It is however noted in Fortney et al. (2008) that the non-equilibrium models should be regarded as being preliminary. Thus, a better understanding of the spectral behavior in this wavelength range might lead to a determination of whether or not the planet is metal-enhanced or not, and thereby provide further clues to its formation.

The results also have implications for imaging surveys for planet detection. The L$^{\prime}$-band range is excellent for planet imaging purposes, and it has been shown in Janson et al. (2009) that a narrower band in the 4 $\mu$m range is even better for bright targets, since it improves the physical contrast according to theoretical models. This Letter provides the first confirmation that this holds true also for a real planet. Evaluating the planet-to-star contrast in the spectral range corresponding to the 4.05 $\mu$m filter gives 8.66 mag, whereas the broad L$^{\prime}$-band contrast is known to be 9.50 mag. Hence, the gain is 0.84 mag for a $\sim$1100 K object, and is expected to increase further for cooler objects.

In the future, further observations with NACO can yield a spectrum also of HR 8799 b and maybe d, and yield a broader coverage HR 8799 c by optimizing the observation procedure, as well as confirming the observed slope. This will provide the opportunity for comparative exoplanetology within a single system. If the models in Fortney et al. (2008) give a correct indication, the L$^{\prime}$M-band range will be important for understanding planetary atmospheres and formation, and future instruments will be available to explore the range more deeply (e.g. NIRSpec on JWST, METIS on E-ELT). Spectroscopy in the shorter wavelength range of JHK-band will likely start to open with dedicated integral field units on VLT (SPHERE) and Gemini (GPI). Integral field units in these bands are available already today, but in the case of SINFONI for instance, the image slicing technique makes it difficult to reconstruct a high-quality PSF, which limits the contrast performance (Janson et al. 2008a), hence planet spectroscopy with such an instrument may turn out to be excessively challenging. In cases where the detection limit is dominated by speckle noise to a greater extent than in the case presented here, exoplanet spectroscopy would be aided by use of spectral deconvolution, as implemented by Thatte et al. (2007) for SINFONI and simulated by Vigan et al. (2008) for future use with, e.g., SPHERE. This technique is based on the fact that the photo-center of a speckle with respect to the primary scales with $\lambda /D$, whereas that of a companion is independent of wavelength, which can be taken advantage of to increase the achievable contrast.

\acknowledgments

We wish the thank the servicing staff at VLT and Subaru for their support, and Jonathan Fortney and Didier Saumon for useful discussion. M.J. is supported through the Reinhardt postdoctoral fellowship from the University of Toronto.



{\it Facilities:} \facility{VLT (NACO)}, \facility{Subaru (IRCS)}.


\clearpage

\end{document}